\documentclass[twocolumn,amsmath,amssymb,aps,pra]{revtex4-1}
\usepackage{graphicx}
\usepackage{dcolumn}
\usepackage{bm}
\usepackage{hyperref}
\usepackage{amsmath} 

\usepackage{color}
\usepackage{textcomp}

\begin{document}

\title[Reconstruction of attosecond pulses trains at ELI-ALPS]{Reconstruction of attosecond pulses in the presence of interfering dressing fields using the 100~kHz ELI-ALPS HR-1 laser system}

\author{D. Hammerland}
\thanks{Equally contributed}
\author{P. Zhang}
\thanks{Equally contributed}
\affiliation{Laboratorium f{\"u}r Physikalische Chemie, ETH Z{\"u}rich, Vladamir Prelog-Weg 2, Z{\"u}rich 8093, Switzerland.}

\author{S. K{\"u}hn}
\author{P. Jojart}
\author{I. Seres}
\author{V. Zuba}
\author{Z. Varallyay}

\author{K. Osvay}
\affiliation{Extreme Light Infrastructre Attosecond Laser Pulse Source, Wolfgang Sandner utca 3, 6728 Szeged, Hungary.}

\author{T. T. Luu}
\email{trung.luu@phys.chem.ethz.ch}

\author{H. J.  W{\"o}rner}
\affiliation{Laboratorium f{\"u}r Physikalische Chemie, ETH Z{\"u}rich, Vladamir Prelog-Weg 2, Z{\"u}rich 8093, Switzerland}

\begin{abstract}
Attosecond Pulse Trains (APT)  generated by high-harmonic generation (HHG) of high-intensity near-infrared (IR) laser pulses have proven valuable for studying the electronic dynamics of atomic  and molecular species. However, the high intensities required for high-photon-energy, high-flux HHG usually limit the class of adequate laser systems to repetition rates below 10~kHz. Here, APT's generated from the 100~kHz, 160~W, 40~fs laser system (HR1) of the Extreme Light Infrastructure Attosecond Light Pulse Source (ELI-ALPS) are reconstructed using the Reconstruction of Attosecond Beating By Interference of two-photon Transitions (RABBIT) technique. These experiments constitute the first attosecond time-resolved photoelectron spectroscopy measurements performed at 100~kHz repetition rate and the first attosecond experiments performed at ELI-ALPS. These RABBIT measurements were taken with an additional IR field temporally locked to the extreme-ultraviolet APT, resulting in an atypical $\omega$ beating. We show that the phase of the $2\omega$ beating recorded under these conditions is strictly identical to that observed in standard RABBIT measurements within second-order perturbation theory. This work highlights an experimental simplification for future experiments based on attosecond interferometry (or RABBIT), which is particularly useful when lasers with high average powers are used.
\end{abstract}

\maketitle


%

\section{Introduction}
Ever since the discovery of high-harmonic generation (HHG) \cite{Lompre1989}, it has proven a valuable source of tabletop extreme-ultraviolet (XUV) radiation. As an added benefit over synchrotrons and X-ray free-electron lasers, HHG can produce laser pulses with attosecond temporal duration \cite{Antoine1996,Paul2001}, making it a powerful tool for studying electronic dynamics on sub-femtosecond time scales \cite{Schultze2010,Klunder2011,Dahlstrom2012,Dahlstrom2012a,Guenot2012,Guenot2014,Palatchi2014,Heuser2016,Huppert2016a,Jordan2017,Loriot2017,Bray2018,Vos2018,Goldsmith2019}. If a many-cycle infrared (IR) pulse is used for HHG, a set of discrete  XUV harmonics are produced, resulting in an attosecond pulse train (APT) \cite{Antoine1996, Paul2001}. 

Characterizing these APT's is challenging as traditional methods of optical gating, such as Frequency Resolved Optical Gating \cite{Trebino1993} and Spectral Phase Interferometry for Direct Electric-field Reconstruction \cite{Iaconis1998}, require passing the pulse through an optical medium, all of which are highly absorptive in the XUV range and thus unsuitable for APT characterization. As a result, the Reconstruction of Attosecond Beating By Interference of two-photon Transitions (RABBIT), was developed in order to characterize APT's \cite{Veniard1996, Paul2001,Muller2002}. In this method, two adjacent harmonics photoionize an atom in the presence of a long-wavelength field, usually the generating IR field, resulting in the creation of side bands. The corresponding side-band intensity beats at twice the frequency of the IR pulse with an offset phase corresponding to the spectral-phase difference of the neighboring harmonics and the atomic phases \cite{Dahlstrom2013}. These atomic phases have been measured experimentally and can be reasonably predicted by theory for atoms \cite{Palatchi2014} and molecules \cite{Baykusheva2017}, allowing for one to obtain solely the XUV spectral phase, which enables the reconstruction of the APT.

A critical aspect of progressing the application of APT's is increasing the repetition rates of the generating IR laser systems. Historically, the peak intensities necessary for high-flux or high-photon-energy HHG, typically $> 10^{14}$ W/cm$^2$, have meant that viable laser systems were limited to amplifiers bottle-necked to less than 10~kHz repetition rates. Although other methods such as enhancement cavities \cite{Jones2005,Pupeza2013,Carstens2016} or tight-focusing geometries with lower pulse power \cite{Higuet2009,Russbueldt2011,Rothhardt2014,Harth2017} have achieved HHG at higher repetition rates, both methods introduce additional experimental challenges that can limit their applicability. The Extreme Light Infrastructure Attosecond Light Pulse Source (ELI-ALPS) facility \cite{Kuhn2017} has been constructed in order to push the frontiers of attosecond science. This facility boasts the High Repetition-1(HR-1) laser system that is designed to eventually produce $>$ 1~mJ pulses with sub-7~fs pulses at 100~kHz, making it the first of a new generation of high-pulse-energy, high-repetition-rate, femtosecond laser systems.

A considerable challenge of these high-average-power systems is the thermal load on the optical components, resulting from even small fractions of the absorbed high average laser power. This is particularly problematic for the few-hundred-nanometer-thick metallic foils that are used to separate the XUV light from its generating IR light. We establish a theoretical framework for RABBIT measurements realized in the presence of a residual generating IR field that is temporally locked to the XUV APT. This IR field interferes with the delayable, dressing IR field and results in a side-band oscillation frequency of $\omega$, the angular frequency of the IR field generating the harmonics, in addition to the traditional $2\omega$ oscillation. We show that the latter encodes exactly the same information as traditional RABBIT measurements performed in the absence of the locked IR field. This method proves to be a valuable extension of the traditional measurement schemes as it captures the same information as traditional RABBIT, but operates without the need for fragile XUV filters. 

Section 2 of this work describes the ELI-ALPS HR-1 laser system, the beamline used for our experiments, and further discusses the challenges of working with high-repetition-rate, high-pulse-energy laser systems. Section 3 focuses on the effect of the temporally locked IR pulse on the RABBIT experiments. Section 4 details a method of reconstruction of the XUV APT and applies it to the XUV field used in this work. Section 5 summarizes the results. 

\section{Experimental setup}
\subsection{ELI-ALPS HR-1 system}
\begin{figure}
    \includegraphics[width=.5\textwidth]{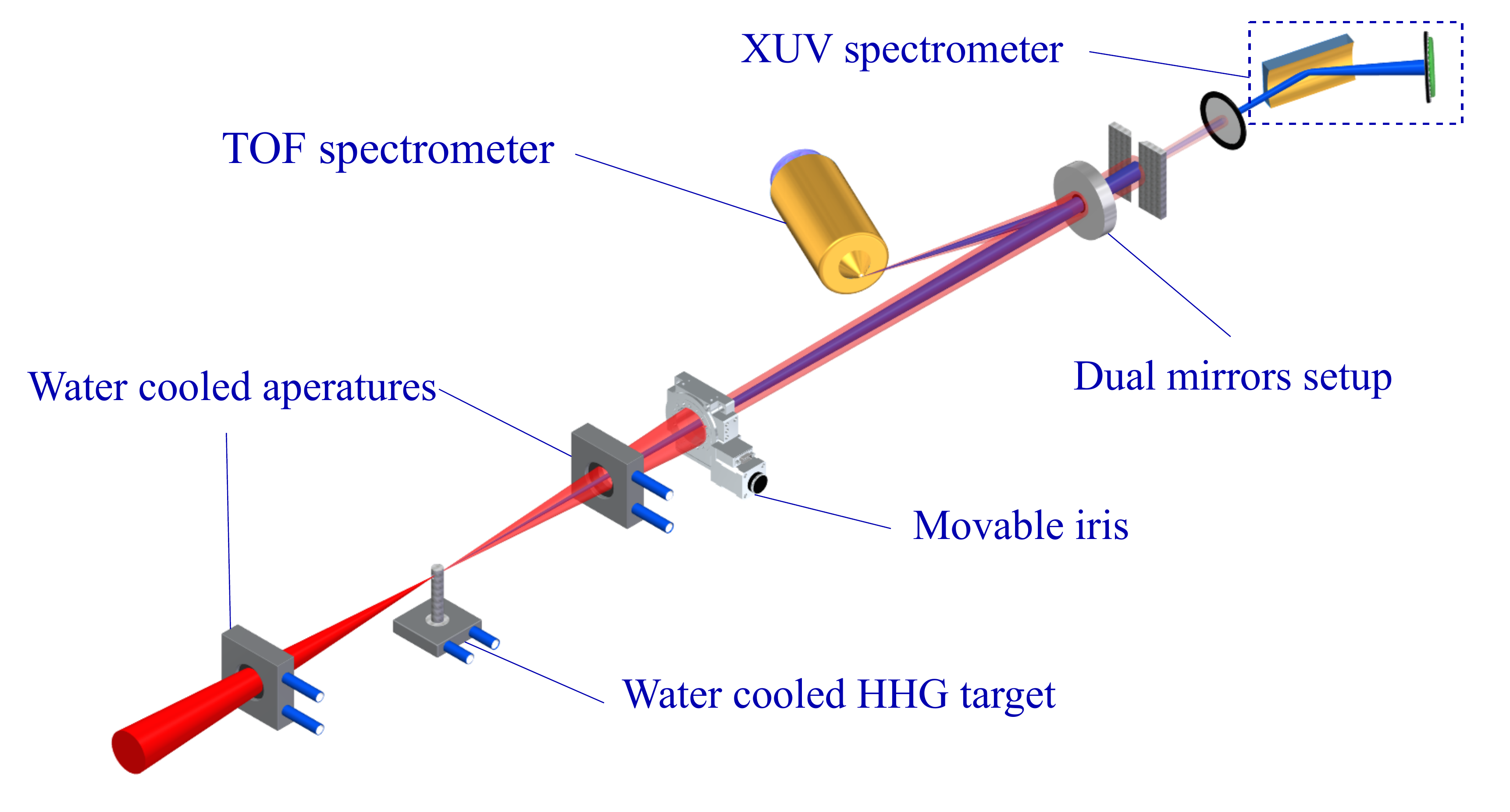}
            \caption{Schematic of the attosecond beamline.}
            \label{attofrontline}
\end{figure} 
The APT was generated using the state-of-the-art ELI ALPS HR-1 laser system. The ytterbium-fiber, chirped-pulse amplifier consists of eight separate amplifiers that are temporally synchronized to output 3~mJ, 200~fs, laser pulses of central wavelength 1030~nm at 100~kHz, summing to an impressive 300 W average power. After the main amplifier, the beam passes through a hollow-core-fiber compression stage where self-phase-modulation broadens the spectrum to support a 40~fs pulse. The 1.6~mJ pulses exiting the fiber are actively pointing stabilized with a beam-pointing system and then diverted through a half-wave-plate and a thin-film polarizer to implement a continuous power adjustment. The beam is expanded using ultraviolet fused silica lenses to a $1/e^2$ diameter of 15 mm and a water-cooled 20 mm aperture. In this configuration, the maximally available average power is around 120 W. Prior to entering the attosecond end station, the pulse duration is compressed to $\sim$40~fs using a chirped mirror compressor.

\subsection{HHG and attosecond beamline}
The attosecond beamline used with the HR-1 laser is shown in Figure \ref{attofrontline}. The laser light was focused for HHG using a 350 mm focal length, 1030~nm high-reflectivity, spherical mirror. The HHG target was a custom-made finite gas-cell chamber with a stainless steel target encapsulated in a differentially pumped cell (excluded for clarity). The metal target was backed with 300-400 mbar of argon. 

After HHG, the emitted harmonics and co-propagating IR fields passed concentrically through an 8 mm hard aperture for heat dissipation and then an iris. This iris was positioned under vacuum using a motorized stage such that the XUV light and the iris center are concentric. The iris opening was then adjusted such that the IR intensity would only induce single-photon transitions during the RABBIT measurements \cite{Swoboda2009}. The co-propagating IR light is helpful in the alignment of the XUV light into an XUV spectrometer. The XUV light is incident at $3^\circ$ onto a variable-line-spacing grating that then disperses the XUV light onto an imaging multi-channel plate (MCP) that emits onto a phosphor screen for detection with a camera. An anodized aluminum blocker is used to prevent the zeroth-order reflection from damaging the MCP.

\begin{figure} 
\centering
    \includegraphics[width=.5\textwidth]{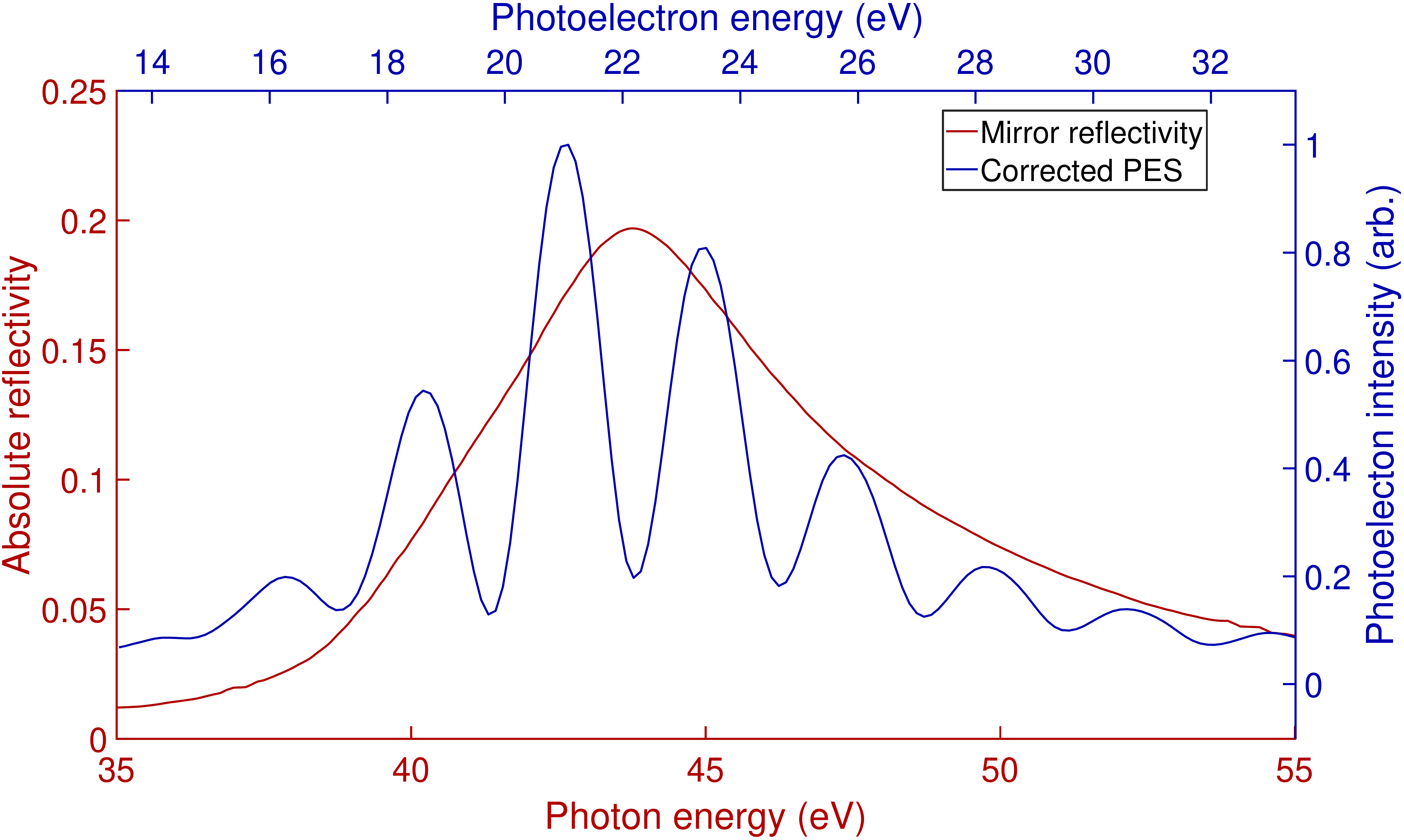}
\caption{Plot of the absolute XUV mirror reflectivity (red) and the cross-section-corrected photoelectron spectrum (blue) (PES). This is obtained by dividing a measured PES by the photoionization cross section at a particular photon energy. The result is the approximate XUV photon spectrum.} 
\label{XUVmirror}
\end{figure} 

After the HHG spectrum has been optimized, a  passively stable dual-mirror assembly is inserted into the beam path for focusing the IR and XUV light onto the target. A piezo-controlled stage is used to adjust the time delay between the XUV and IR pulses. The inner mirror was a custom-made XUV multi-layer mirror (AXO Dresden GmbH) with a reflectivity designed to be centered at $\sim$43~eV and a $\sim$7~eV bandwidth, as shown in Figure \ref{XUVmirror}. By dividing the a static PES from neon by the photoionization cross section, as shown in Figure \ref{XUVmirror}, the XUV spectra can be approximated. This demonstrates how the mirror allows for the isolation of a small group of XUV harmonics. This XUV mirror has a radius of curvature (ROC) of 0.500~m. The outer mirror was a high-reflectivity, 1030~nm optic, also with a ROC of 0.500~m. The XUV and IR light were then focused in front of a 0.500~m field-free time-of-flight (TOF) spectrometer. This apparatus has been described in detail in \cite{Jordan2017}. 

After the light passes through the sample, it is guided towards an imaging setup. This allows for fine positioning of the sample target and beam focus relative to the TOF skimmer. Further, it can be used to monitor the IR beam positions and focal parameters.  

\subsection{100 kHz considerations}\label{100 kHz considerations}
Two major challenges of using high-intensity, high-repetition-rate laser systems are managing the high thermal load on the optics and the high speed of data acquisition. When working with an average power output exceeding 100~W, even a 99$\%$ reflectivity results in over 1~W of optical power being absorbed, enough to cause wavefront distortion from local heating of the optics. As a result, it was critical to use highly reflective, $>$ 99.9 $\%$, 1030~nm optics for all guiding optics and for focusing into the HHG chamber and interaction region of the TOF spectrometer. 

An additional problem that arose was the inability to use metallic foils for filtering the XUV light from the generating IR light. These 200-nm aluminum foils (Lebow Company) burnt at incident-power levels of only 10 W. The theoretical implications of this additional temporally locked IR field on the measured data are addressed in Section 4. An additional challenge arising from the impossibility of using filters is that the XUV light co-propagates onto the XUV mirror with a portion of its generating IR field, preventing their separation. However, the imaging setup did not reveal notable thermally induced deformation of the XUV mirror, as seen by the stability of the IR beam profile during operation. Additionally, the PES was constant over these times, so that the XUV beam distortion also appeared to be negligible. Thermal expansion of the XUV mirror was, however, observed. The temporal distortion of the measurement was prevented by allowing the XUV mirror to thermalize before scans. 

Wherever beam attenuation was needed considerable efforts were made to dissipate heat actively in order to optimize stability. Custom water-cooled elements were added throughout the system, as shown in Figure \ref{attofrontline}. The 15 mm hard aperture before the chirped mirror compressor and the post-HHG, 8 mm, hard aperture are both connected to a closed, chilled water loop that was used to dissipate approximately 20 W and 35 W, respectively. The HHG target featured a similar cooling system. This increased the stability of the HHG and mitigated thermal damage to the HHG target. 

In order to record the time-of-flight signals at high repetition rate, we employed a high-speed digitizer (Keysight Technologies U5309A PCIe, 2 GS/s, 2 channels, 8 bits) directly connected to the MCP output. Although Keysight Technologies could provide the firmware option to do on-the-fly averaging, we found that the triggered simultaneous acquisition and readout mode already supports real-time acquisition at 100~kHz. By making use of the ring buffer, optimal direct memory access transfer can be obtained such that each data transfer is very close to one megabyte per transfer. In this case, a LabVIEW virtual instrument can be written to simultaneously record and transfer the data from the digitizer to the computer in parallel. As a result, we can record the full time-of-flight spectrum for every laser shot using the digitizer and average the signals on the CPU in real time at 100~kHz for time-of-flight intervals of 400 ns. This is sufficient for all of the measurements presented in this work. For longer time-of-flight intervals the actual acquisition speed drops but even for extremely long intervals, the achievable recording speed hardly falls below 50~kHz. By repeating the data acquisition and averaging for every time delay, we obtained the RABBIT traces, making use of all of the laser shots at 100~kHz.

\section{Results}\label{w2w}
As mentioned in Section \ref{100 kHz considerations}, a notable challenge of high-average-power HHG experiments comes when attempting to isolate the XUV photons from their generating IR light. Few-hundred-nanometer-thick metallic foils have been historically used to remove the IR light and compensate for the attochirp. These foils, however, are notoriously fragile and cannot handle the high average power present in high-repetition-rate, high-pulse-energy laser systems. Annular beams can be used to generate spatially separated XUV and IR pulses \cite{Gaumnitz2018}, but masks used to block out the central areas of the generating beam also suffer from damage-threshold issues with high-average-power systems and drilled mirrors are costly and sacrifice generating power required for the HHG process. 

In this work, a new method is introduced and demonstrated wherein the IR light was not separated from the XUV light, but rather used in combination with the delayed IR light. This additional laser field can be easily incorporated in the common RABBIT framework by introducing a second IR field that is temporally locked to the XUV APT. Adding this field and following through the standard RABBIT derivations (see appendix), the typical RABBIT sideband oscillations \cite{Dahlstrom2013}, 
\begin{eqnarray}\label{twow}
    SB(\tau) \propto \cos\big(2 \omega\tau  + \Delta\phi_{\mathrm{XUV}} + \Delta\phi_{\mathrm{atomic}}\big),
\end{eqnarray}
are still preserved, although they are additionally accompanied by oscillations at angular frequency $\omega$. Nevertheless, the phases of the 2$\omega$ oscillations are strictly identical to the situation encountered in traditional RABBIT experiments. In Eq. (\ref{twow}) $\Delta\phi_{\mathrm{XUV}}$ represents the spectral-phase difference between the next-higher and next-lower harmonic orders, whereas $\Delta\phi_{\mathrm{atomic}}$ represents the corresponding atomic-phase difference.

\begin{figure} \label{RABBIT}
    \includegraphics[width=.45\textwidth]{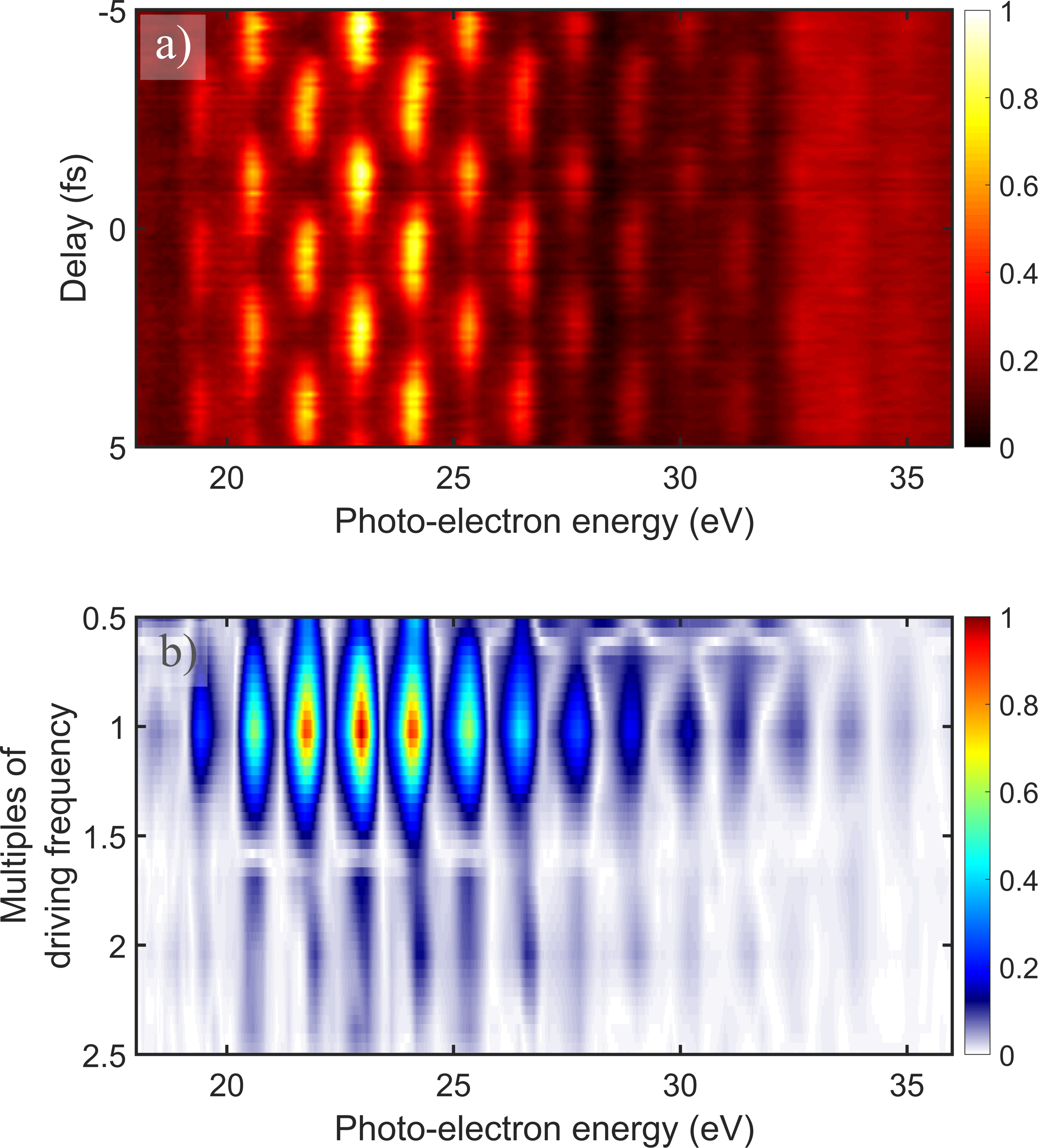}
            \caption{False-color representation of a) the attosecond PES scan, b) the Fourier transform amplitude of the scan along the time axis, showing oscillations with angular frequencies of both $\omega$ and $2\omega$. }
            \label{PES}
\end{figure} 

The relative intensity of the $\omega$- and $2\omega$- oscillations is proportional to the relative intensity of the delayed and locked IR fields. In the limit of a zero-intensity locked IR field, the standard RABBIT scheme is recovered and the $2\omega$ oscillations dominate. 
When the delayed IR is of zero intensity, a static spectrum is expected. These limits were observed during the implementation of the experiment. 


Figure \ref{RABBIT}a) shows an attosecond PES trace measured in neon. The 1030~nm generating field has an optical period of 3.4~fs, which dominates the oscillations. However, Fourier transformation along the temporal-delay axis, shown in Figure \ref{RABBIT}b), reveals that weaker $2\omega$ oscillations are also present in the data. This places the measurement in the limit of a weak delayed IR field relative to the locked IR field.  

\section{Pulse reconstruction}\label{Reconstruction}
\begin{figure}
    \includegraphics[width=.5\textwidth]{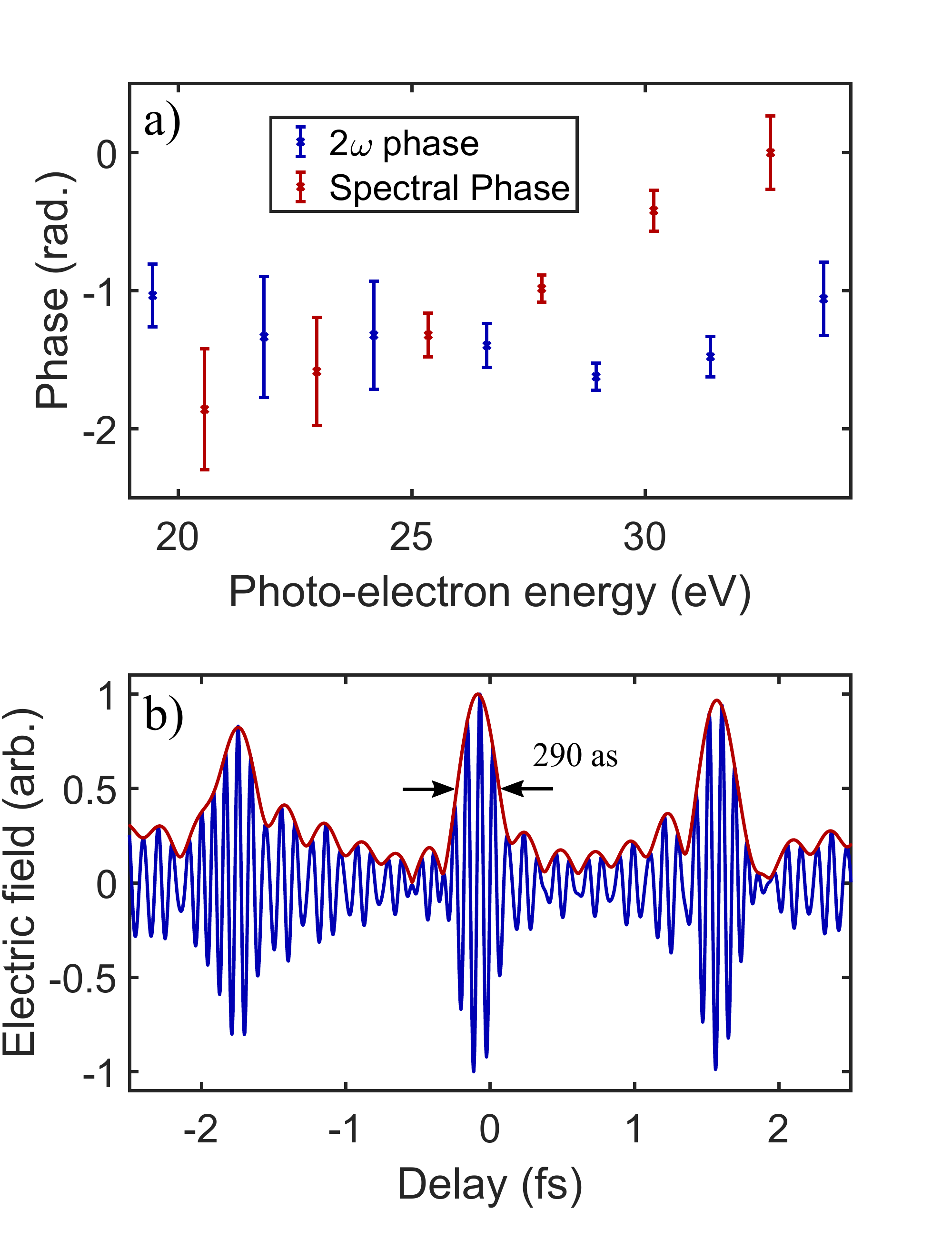}
        \caption{Reconstruction of the APT. a) Phase extracted from the $2\omega$ oscillations shown in Figure \ref{PES} and reconstructed spectral phase, after taking into account the atomic phases, b) attosecond structure of the generated APT (blue) and its envelope (red). $\tau$ represents the full width at half maximum (FWHM) of the electric-field envelope of $\sim$290~as, corresponding to an intensity FWHM of $\sim$205~as. }
\label{NeReconstruction}
\end{figure}

The phase of the $2\omega$ oscillations in each sideband were extracted by taking the angle of the complex average of the Fourier transform across the photoelectron-energy bandwidth of each sideband. The phases of the sidebands are plotted (blue) in Figure \ref{NeReconstruction}a). From each of these values, the difference of the atomic phases of the neighboring harmonics taken from \cite{Palatchi2014} were subtracted to determine the XUV spectral phase, $\Delta\phi_{XUV}$, by concatenation. More precisely, the highest-energy sideband was chosen for the zero-phase value in the reconstruction and the relative phases of the harmonics were then found by successively adding the phase difference from one harmonic to the next. We note that the determination of accurate error bars from these measurements represents a significant challenge, as instability in gas pressure, laser fluctuations, mirror assembly temperature and stability all play a significant role but cannot be recorded on a shot-to-shot basis. Here they were approximated by adding in quadrature the scan-normalized signal-to-noise ratios with the standard deviation of the weighted phase values.

The spectral intensity of each harmonic was produced by dividing the PES by the associated neon 2p photoionization cross section \cite{Becker1996}. Delta functions at the XUV harmonic energy were assumed to approximate the APT near the center of the pulse. These fields were superimposed with their corresponding relative phases and the intensities in order to reconstruct the APT, Figure \ref{NeReconstruction}b) (blue), near the center of the pulse. The CEP of the XUV APT cannot be determined within the RABBIT approach, but the envelope (red) can be shown with higher certainty.


\section{Conclusion and outlook}
This paper describes the first attosecond time-resolved experiments realized in the ELI-ALPS facility and marks the first demonstration of 100~kHz attosecond PES, representing a substantial step forward in the development of table-top attosecond experiments. These higher acquisition rates open up exciting avenues for emission-angle-resolved attosecond photoelectron spectroscopy in all phases of matter and will be particularly valuable for electron-ion-coincidence type measurements that currently remain repetition-rate limited. 

The RABBIT technique has been tailored to work with an additional IR pulse temporally locked to the XUV APT. The manifestation of this pulse in the PES is an $\omega$ oscillation in addition to the usual $2\omega$ oscillation. Theoretically, the $2\omega$ oscillations were shown to encode identical information as in the standard RABBIT scheme. This opens the possibility of performing RABBIT measurements without the need for fragile filters and will thereby benefit attosecond pulse metrology as well as measurements of attosecond photoionization delays at high repetition rates and high average powers. 

\section{Acknowledgement}
The authors acknowledge the contributions of Dr. Arohi Jain and Dr. Thomas Gaumnitz to the preparatory phase of this work, such as the installation of the attosecond beamline at the ELI-ALPS facility, as well as optical and mechanical adaptations for the operation at high laser repetition rates. The authors further acknowledge all of the staff at the ELI-ALPS laser facility for their efforts throughout the beam times, particularly Dr. Tam{\'a}s Csizmadia and Dr. Mikl{\'o}s F{\"u}le, the technical staff of the ETH Z{\"u}rich Laboratorium f{\"u}r Physikalische Chemie, particularly Andreas Schneider for help with electronics as well as Mario Seiler and Markus Kerellaj for the custom-manufactured mechanical parts used during the experiment and Dr. Adam Smith, Danylo Matselyukh, and Jakub Koc{\'a}k for fruitful discussions and proofreading. 

This research was funded by ETH Z{\"u}rich and the Swiss National Science Foundation through grants 200021E\_162822 and 200021\_172946.   

\appendix
\section*{Appendix}
\setcounter{section}{1}

Here, we derive the expressions for the side-band-intensity oscillations relevant for the interpretation of the present experimental results. For simplicity, we restrict the description of the XUV APT to that of two neighboring harmonics of orders $q+1$ and $q-1$
\begin{multline} \label{EXUV} 
E_{\mathrm{XUV}}(t) = \Bigg( A_{q+1} e^{i\big((q+1)\omega t + \phi_{q+1}\big)}  + \\
A_{q-1} e^{i\big((q-1) \omega t+ \phi_{q-1} \big)} + C.C. \Bigg),
\end{multline}
where $\phi_{q+1}$ and $\phi_{q-1}$ are the relevant spectral-phase components and $C.C.$ stands for complex conjugate.
We further describe the IR field as consisting of two components, one that is temporally locked to the XUV APT and one that is delayable (with delay $\tau$) with respect to it:
\begin{multline} \label{EIRinANDout} \scriptsize
E_{IR}(\tau,\omega)= \Bigg( A_\mathrm{{locked}} e^{i(\omega t+ \phi_{\mathrm{rel}})}  + A_\mathrm{{locked}}^* e^{-i(\omega t+ \phi_{\mathrm{rel}})} \\ + 
A_\mathrm{{delayed}} e^{i\omega (t-\tau )} + A_\mathrm{{delayed}}^* e^{-i\omega(t-\tau )}\Bigg).
\end{multline}
Following the standard framework of second-order perturbation theory of RABBIT, the induced polarization of the combined XUV and IR pulses describing two-color two-photon ionization of an atom can be written as
\begin{multline} \label{RABBITPLIRpolarizationfields}
\begin{split}
P(\tau,\omega) \propto  \Bigg( A_\mathrm{{locked}} e^{i(\omega t+ \phi_{\mathrm{rel}})}+ A_\mathrm{{locked}}^* e^{-i(\omega t+ \phi_{\mathrm{rel}})} \\ 
+ A_\mathrm{{delayed}} e^{i\omega (t-\tau )} + A_\mathrm{{delayed}}^* e^{-i\omega(t-\tau )}\Bigg) \\
\times \Bigg( A_{q+1} e^{i\big((q+1)\omega t + \phi_{q+1} + \phi^\mathrm{at}_{q+1}\big)} \\
+ A_{q-1} e^{i\big((q-1) \omega t+ \phi_{q-1} + \phi^\mathrm{at}_{q-1}}\big) + C.C. \Bigg),
\end{split}
\end{multline}
\noindent where $\phi^\mathrm{at}_{q+1}$ and $\phi^\mathrm{at}_{q-1}$ are the relevant atomic phases resulting from two-color two-photon ionization. Hence, the result is merely the superposition of two RABBIT-like polarizations, one generated from the locked IR field and one generated from the delayed IR field:
\begin{multline} \label{RABBITPLIRpolarizationtot}
\begin{split}
P(\tau,\omega) & \propto A_\mathrm{{delayed}}A_{q+1} e ^{i\big((q+2)\omega t - \omega \tau + \phi_{q+1} + \phi^\mathrm{at}_{q+1} \big)} \\
& + A_\mathrm{{delayed}}^*A_{q+1} e ^{i(q\omega t + \omega \tau + \phi_{q+1} + \phi^\mathrm{at}_{q+1}) } \\
& + A_\mathrm{{delayed}}A_{q-1} e ^{i(q\omega t - \omega \tau + \phi_{q-1} + \phi^\mathrm{at}_{q-1})} \\
& + A_\mathrm{{delayed}}^*A_{q-1} e ^{i\big((q-2)\omega t + \omega \tau + \phi_{q-1} + \phi^\mathrm{at}_{q-1}\big) } \\
& + A_\mathrm{{locked}}A_{q+1} e ^{i\big((q+2)\omega t + \phi_{\mathrm{rel}} + \phi_{q+1} + \phi^\mathrm{at}_{q+1} \big)} \\
& + A_\mathrm{{locked}}^*A_{q+1} e ^{i(q\omega t - \phi_{\mathrm{rel}} + \phi_{q+1} + \phi^\mathrm{at}_{q+1}) } \\
& + A_\mathrm{{locked}}A_{q-1} e ^{i(q\omega t + \phi_{\mathrm{rel}} + \phi_{q-1} + \phi^\mathrm{at}_{q-1})} \\
& + A_\mathrm{{locked}}^*A_{q-1} e ^{i\big((q-2)\omega t - \phi_{\mathrm{rel}} + \phi_{q-1} + \phi^\mathrm{at}_{q-1}\big) }\\
& + C.C. 
\end{split}
\end{multline}


\noindent Examining the polarization at frequency $q\omega$, corresponding to the side-band oscillation we find 
\begin{multline} \label{RABBITPLIRpolarizationq}
\begin{split}
P_q(\tau,\omega) \propto A_\mathrm{{delayed}}^*A_{q+1} e ^{i(q\omega t + \omega \tau + \phi_{q+1} + \phi^\mathrm{at}_{q+1}) } \\
+ A_\mathrm{{delayed}}A_{q-1} e ^{i(q\omega t - \omega \tau + \phi_{q-1} +  \phi^\mathrm{at}_{q-1})}  \\
+ A_\mathrm{{locked}}^*A_{q+1} e ^{i(q\omega t - \phi_{\mathrm{rel}} + \phi_{q+1} + \phi^\mathrm{at}_{q+1})} \\
+ A_\mathrm{{locked}}A_{q-1} e ^{i(q\omega t + \phi_{\mathrm{rel}} + \phi_{q-1} + \phi^\mathrm{at}_{q-1})}.
\end{split}
\end{multline}

\noindent The intensity oscillation at the relevant $2\omega$ frequency is
\begin{multline} \label{RABBITPLIRfreqq+2}
\begin{split}
&S_{q,2\omega}(\tau) \propto \left|P_{q,2\omega}(\tau,\omega)\right|^2 \propto \\
&2+2\cos(2\omega\tau + \phi_{q+1} - \phi_{q-1} + \phi^{\mathrm{at}}_{q+1} - \phi^\mathrm{at}_{q-1}), 
\end{split}
\end{multline}
which is identical with the standard RABBIT situation and is reproduced as Eq. (1) in the main text, with the correspondence $\Delta\phi_{\rm XUV}=\phi_{q+1}-\phi_{q-1}$ and $\Delta\phi_{\rm atomic}=\phi^{\rm at}_{q+1}-\phi^{\rm at}_{q-1}$.


\bibliography{ELIw2w}

\begin{thebibliography}{34}%
\makeatletter
\providecommand \@ifxundefined [1]{%
 \@ifx{#1\undefined}
}%
\providecommand \@ifnum [1]{%
 \ifnum #1\expandafter \@firstoftwo
 \else \expandafter \@secondoftwo
 \fi
}%
\providecommand \@ifx [1]{%
 \ifx #1\expandafter \@firstoftwo
 \else \expandafter \@secondoftwo
 \fi
}%
\providecommand \natexlab [1]{#1}%
\providecommand \enquote  [1]{``#1''}%
\providecommand \bibnamefont  [1]{#1}%
\providecommand \bibfnamefont [1]{#1}%
\providecommand \citenamefont [1]{#1}%
\providecommand \href@noop [0]{\@secondoftwo}%
\providecommand \href [0]{\begingroup \@sanitize@url \@href}%
\providecommand \@href[1]{\@@startlink{#1}\@@href}%
\providecommand \@@href[1]{\endgroup#1\@@endlink}%
\providecommand \@sanitize@url [0]{\catcode `\\12\catcode `\$12\catcode
  `\&12\catcode `\#12\catcode `\^12\catcode `\_12\catcode `\%12\relax}%
\providecommand \@@startlink[1]{}%
\providecommand \@@endlink[0]{}%
\providecommand \url  [0]{\begingroup\@sanitize@url \@url }%
\providecommand \@url [1]{\endgroup\@href {#1}{\urlprefix }}%
\providecommand \urlprefix  [0]{URL }%
\providecommand \Eprint [0]{\href }%
\providecommand \doibase [0]{https://doi.org/}%
\providecommand \selectlanguage [0]{\@gobble}%
\providecommand \bibinfo  [0]{\@secondoftwo}%
\providecommand \bibfield  [0]{\@secondoftwo}%
\providecommand \translation [1]{[#1]}%
\providecommand \BibitemOpen [0]{}%
\providecommand \bibitemStop [0]{}%
\providecommand \bibitemNoStop [0]{.\EOS\space}%
\providecommand \EOS [0]{\spacefactor3000\relax}%
\providecommand \BibitemShut  [1]{\csname bibitem#1\endcsname}%
\let\auto@bib@innerbib\@empty
\bibitem [{\citenamefont {Lompr{\'{e}}}\ \emph {et~al.}(1989)\citenamefont
  {Lompr{\'{e}}}, \citenamefont {L'Huillier},\ and\ \citenamefont
  {Mainfray}}]{Lompre1989}%
  \BibitemOpen
  \bibfield  {author} {\bibinfo {author} {\bibfnamefont {L.~A.}\ \bibnamefont
  {Lompr{\'{e}}}}, \bibinfo {author} {\bibfnamefont {A.}~\bibnamefont
  {L'Huillier}},\ and\ \bibinfo {author} {\bibfnamefont {G.}~\bibnamefont
  {Mainfray}},\ }\bibfield  {title} {\bibinfo {title} {{Harmonic generation in
  rare gases at high laser intensity}},\ }\href
  {https://doi.org/10.1007/3-540-51430-9_6} {\bibfield  {journal} {\bibinfo
  {journal} {Fundamentals of Laser Interactions II}\ }\textbf {\bibinfo
  {volume} {39}},\ \bibinfo {pages} {67} (\bibinfo {year} {1989})}\BibitemShut
  {NoStop}%
\bibitem [{\citenamefont {Antoine}\ \emph {et~al.}(1996)\citenamefont
  {Antoine}, \citenamefont {L'huillier},\ and\ \citenamefont
  {Lewenstein}}]{Antoine1996}%
  \BibitemOpen
  \bibfield  {author} {\bibinfo {author} {\bibfnamefont {P.}~\bibnamefont
  {Antoine}}, \bibinfo {author} {\bibfnamefont {A.}~\bibnamefont
  {L'huillier}},\ and\ \bibinfo {author} {\bibfnamefont {M.}~\bibnamefont
  {Lewenstein}},\ }\bibfield  {title} {\bibinfo {title} {{Attosecond pulse
  trains using high–order harmonics}},\ }\href
  {https://doi.org/10.1103/PhysRevLett.77.1234} {\bibfield  {journal} {\bibinfo
   {journal} {Physical Review Letters}\ }\textbf {\bibinfo {volume} {77}},\
  \bibinfo {pages} {1234} (\bibinfo {year} {1996})}\BibitemShut {NoStop}%
\bibitem [{\citenamefont {Paul}\ \emph {et~al.}(2001)\citenamefont {Paul},
  \citenamefont {M.}, \citenamefont {Toma}, \citenamefont {S.}, \citenamefont
  {Breger}, \citenamefont {P.}, \citenamefont {Mullot}, \citenamefont {G.},
  \citenamefont {Auge}, \citenamefont {F.}, \citenamefont {Balcou},
  \citenamefont {Muller}, \citenamefont {G.},\ and\ \citenamefont
  {Agostini}}]{Paul2001}%
  \BibitemOpen
  \bibfield  {author} {\bibinfo {author} {\bibnamefont {Paul}}, \bibinfo
  {author} {\bibfnamefont {P.}~\bibnamefont {M.}}, \bibinfo {author}
  {\bibnamefont {Toma}}, \bibinfo {author} {\bibfnamefont {E.}~\bibnamefont
  {S.}}, \bibinfo {author} {\bibnamefont {Breger}}, \bibinfo {author}
  {\bibnamefont {P.}}, \bibinfo {author} {\bibnamefont {Mullot}}, \bibinfo
  {author} {\bibnamefont {G.}}, \bibinfo {author} {\bibnamefont {Auge}},
  \bibinfo {author} {\bibnamefont {F.}}, \bibinfo {author} {\bibnamefont
  {Balcou}}, \bibinfo {author} {\bibnamefont {Muller}}, \bibinfo {author}
  {\bibfnamefont {H.}~\bibnamefont {G.}},\ and\ \bibinfo {author} {\bibnamefont
  {Agostini}},\ }\bibfield  {title} {\bibinfo {title} {{Observation of a Train
  of Attosecond Pulses from High Harmonic Generation}},\ }\href
  {https://doi.org/10.1126/science.1059413} {\bibfield  {journal} {\bibinfo
  {journal} {Science}\ }\textbf {\bibinfo {volume} {292}},\ \bibinfo {pages}
  {1689} (\bibinfo {year} {2001})}\BibitemShut {NoStop}%
\bibitem [{\citenamefont {Schultze}\ \emph {et~al.}(2010)\citenamefont
  {Schultze}, \citenamefont {Karpowicz}, \citenamefont {Gagnon}, \citenamefont
  {Korbman}, \citenamefont {Hofstetter}, \citenamefont {Neppl}, \citenamefont
  {Cavalieri}, \citenamefont {Komninos}, \citenamefont {Nicolaides},
  \citenamefont {Pazourek}, \citenamefont {Nagele}, \citenamefont {Feist},
  \citenamefont {Azzeer}, \citenamefont {Ernstorfer}, \citenamefont
  {Kienberger}, \citenamefont {Kleineberg}, \citenamefont {Goulielmakis},
  \citenamefont {Krausz},\ and\ \citenamefont {Yakovlev}}]{Schultze2010}%
  \BibitemOpen
  \bibfield  {author} {\bibinfo {author} {\bibfnamefont {M.}~\bibnamefont
  {Schultze}}, \bibinfo {author} {\bibfnamefont {N.}~\bibnamefont {Karpowicz}},
  \bibinfo {author} {\bibfnamefont {J.}~\bibnamefont {Gagnon}}, \bibinfo
  {author} {\bibfnamefont {M.}~\bibnamefont {Korbman}}, \bibinfo {author}
  {\bibfnamefont {M.}~\bibnamefont {Hofstetter}}, \bibinfo {author}
  {\bibfnamefont {S.}~\bibnamefont {Neppl}}, \bibinfo {author} {\bibfnamefont
  {A.~L.}\ \bibnamefont {Cavalieri}}, \bibinfo {author} {\bibfnamefont
  {Y.}~\bibnamefont {Komninos}}, \bibinfo {author} {\bibfnamefont {C.~A.}\
  \bibnamefont {Nicolaides}}, \bibinfo {author} {\bibfnamefont
  {R.}~\bibnamefont {Pazourek}}, \bibinfo {author} {\bibfnamefont
  {S.}~\bibnamefont {Nagele}}, \bibinfo {author} {\bibfnamefont
  {J.}~\bibnamefont {Feist}}, \bibinfo {author} {\bibfnamefont {A.~M.}\
  \bibnamefont {Azzeer}}, \bibinfo {author} {\bibfnamefont {R.}~\bibnamefont
  {Ernstorfer}}, \bibinfo {author} {\bibfnamefont {R.}~\bibnamefont
  {Kienberger}}, \bibinfo {author} {\bibfnamefont {U.}~\bibnamefont
  {Kleineberg}}, \bibinfo {author} {\bibfnamefont {E.}~\bibnamefont
  {Goulielmakis}}, \bibinfo {author} {\bibfnamefont {F.}~\bibnamefont
  {Krausz}},\ and\ \bibinfo {author} {\bibfnamefont {V.~S.}\ \bibnamefont
  {Yakovlev}},\ }\bibfield  {title} {\bibinfo {title} {{Delay in
  Photoemission}},\ }\href@noop {} {\bibfield  {journal} {\bibinfo  {journal}
  {Science}\ }\textbf {\bibinfo {volume} {328}},\ \bibinfo {pages} {1658}
  (\bibinfo {year} {2010})}\BibitemShut {NoStop}%
\bibitem [{\citenamefont {Kl{\"{u}}nder}\ \emph {et~al.}(2011)\citenamefont
  {Kl{\"{u}}nder}, \citenamefont {Dahlstr{\"{o}}m}, \citenamefont
  {Gisselbrecht}, \citenamefont {Fordell}, \citenamefont {Swoboda},
  \citenamefont {Gu{\'{e}}not}, \citenamefont {Johnsson}, \citenamefont
  {Caillat}, \citenamefont {Mauritsson}, \citenamefont {Maquet}, \citenamefont
  {Ta{\"{i}}eb},\ and\ \citenamefont {L'Huillier}}]{Klunder2011}%
  \BibitemOpen
  \bibfield  {author} {\bibinfo {author} {\bibfnamefont {K.}~\bibnamefont
  {Kl{\"{u}}nder}}, \bibinfo {author} {\bibfnamefont {J.~M.}\ \bibnamefont
  {Dahlstr{\"{o}}m}}, \bibinfo {author} {\bibfnamefont {M.}~\bibnamefont
  {Gisselbrecht}}, \bibinfo {author} {\bibfnamefont {T.}~\bibnamefont
  {Fordell}}, \bibinfo {author} {\bibfnamefont {M.}~\bibnamefont {Swoboda}},
  \bibinfo {author} {\bibfnamefont {D.}~\bibnamefont {Gu{\'{e}}not}}, \bibinfo
  {author} {\bibfnamefont {P.}~\bibnamefont {Johnsson}}, \bibinfo {author}
  {\bibfnamefont {J.}~\bibnamefont {Caillat}}, \bibinfo {author} {\bibfnamefont
  {J.}~\bibnamefont {Mauritsson}}, \bibinfo {author} {\bibfnamefont
  {A.}~\bibnamefont {Maquet}}, \bibinfo {author} {\bibfnamefont
  {R.}~\bibnamefont {Ta{\"{i}}eb}},\ and\ \bibinfo {author} {\bibfnamefont
  {A.}~\bibnamefont {L'Huillier}},\ }\bibfield  {title} {\bibinfo {title}
  {{Probing single-photon ionization on the attosecond time scale}},\ }\href
  {https://doi.org/10.1103/PhysRevLett.106.143002} {\bibfield  {journal}
  {\bibinfo  {journal} {Physical Review Letters}\ }\textbf {\bibinfo {volume}
  {106}},\ \bibinfo {pages} {1} (\bibinfo {year} {2011})}\BibitemShut {NoStop}%
\bibitem [{\citenamefont {Dahlstr{\"{o}}m}\ \emph
  {et~al.}(2012{\natexlab{a}})\citenamefont {Dahlstr{\"{o}}m}, \citenamefont
  {L'Huillier},\ and\ \citenamefont {Maquet}}]{Dahlstrom2012}%
  \BibitemOpen
  \bibfield  {author} {\bibinfo {author} {\bibfnamefont {J.~M.}\ \bibnamefont
  {Dahlstr{\"{o}}m}}, \bibinfo {author} {\bibfnamefont {A.}~\bibnamefont
  {L'Huillier}},\ and\ \bibinfo {author} {\bibfnamefont {A.}~\bibnamefont
  {Maquet}},\ }\bibfield  {title} {\bibinfo {title} {{Introduction to
  attosecond delays in photoionization}},\ }\bibfield  {journal} {\bibinfo
  {journal} {Journal of Physics B: Atomic, Molecular and Optical Physics}\
  }\textbf {\bibinfo {volume} {45}},\ \href
  {https://doi.org/10.1088/0953-4075/45/18/183001}
  {10.1088/0953-4075/45/18/183001} (\bibinfo {year}
  {2012}{\natexlab{a}})\BibitemShut {NoStop}%
\bibitem [{\citenamefont {Dahlstr{\"{o}}m}\ \emph
  {et~al.}(2012{\natexlab{b}})\citenamefont {Dahlstr{\"{o}}m}, \citenamefont
  {Carette},\ and\ \citenamefont {Lindroth}}]{Dahlstrom2012a}%
  \BibitemOpen
  \bibfield  {author} {\bibinfo {author} {\bibfnamefont {J.~M.}\ \bibnamefont
  {Dahlstr{\"{o}}m}}, \bibinfo {author} {\bibfnamefont {T.}~\bibnamefont
  {Carette}},\ and\ \bibinfo {author} {\bibfnamefont {E.}~\bibnamefont
  {Lindroth}},\ }\bibfield  {title} {\bibinfo {title} {{Diagrammatic approach
  to attosecond delays in photoionization}},\ }\href
  {https://doi.org/10.1103/PhysRevA.86.061402} {\bibfield  {journal} {\bibinfo
  {journal} {Physical Review A - Atomic, Molecular, and Optical Physics}\
  }\textbf {\bibinfo {volume} {86}},\ \bibinfo {pages} {1} (\bibinfo {year}
  {2012}{\natexlab{b}})}\BibitemShut {NoStop}%
\bibitem [{\citenamefont {Gu{\'{e}}not}\ \emph {et~al.}(2012)\citenamefont
  {Gu{\'{e}}not}, \citenamefont {Kl{\"{u}}nder}, \citenamefont {Arnold},
  \citenamefont {Kroon}, \citenamefont {Dahlstr{\"{o}}m}, \citenamefont
  {Miranda}, \citenamefont {Fordell}, \citenamefont {Gisselbrecht},
  \citenamefont {Johnsson}, \citenamefont {Mauritsson}, \citenamefont
  {Lindroth}, \citenamefont {Maquet}, \citenamefont {Ta{\"{i}}eb},
  \citenamefont {L'Huillier},\ and\ \citenamefont {Kheifets}}]{Guenot2012}%
  \BibitemOpen
  \bibfield  {author} {\bibinfo {author} {\bibfnamefont {D.}~\bibnamefont
  {Gu{\'{e}}not}}, \bibinfo {author} {\bibfnamefont {K.}~\bibnamefont
  {Kl{\"{u}}nder}}, \bibinfo {author} {\bibfnamefont {C.~L.}\ \bibnamefont
  {Arnold}}, \bibinfo {author} {\bibfnamefont {D.}~\bibnamefont {Kroon}},
  \bibinfo {author} {\bibfnamefont {J.~M.}\ \bibnamefont {Dahlstr{\"{o}}m}},
  \bibinfo {author} {\bibfnamefont {M.}~\bibnamefont {Miranda}}, \bibinfo
  {author} {\bibfnamefont {T.}~\bibnamefont {Fordell}}, \bibinfo {author}
  {\bibfnamefont {M.}~\bibnamefont {Gisselbrecht}}, \bibinfo {author}
  {\bibfnamefont {P.}~\bibnamefont {Johnsson}}, \bibinfo {author}
  {\bibfnamefont {J.}~\bibnamefont {Mauritsson}}, \bibinfo {author}
  {\bibfnamefont {E.}~\bibnamefont {Lindroth}}, \bibinfo {author}
  {\bibfnamefont {A.}~\bibnamefont {Maquet}}, \bibinfo {author} {\bibfnamefont
  {R.}~\bibnamefont {Ta{\"{i}}eb}}, \bibinfo {author} {\bibfnamefont
  {A.}~\bibnamefont {L'Huillier}},\ and\ \bibinfo {author} {\bibfnamefont
  {A.~S.}\ \bibnamefont {Kheifets}},\ }\bibfield  {title} {\bibinfo {title}
  {{Photoemission-time-delay measurements and calculations close to the
  3s-ionization-cross-section minimum in Ar}},\ }\href
  {https://doi.org/10.1103/PhysRevA.85.053424} {\bibfield  {journal} {\bibinfo
  {journal} {Physical Review A - Atomic, Molecular, and Optical Physics}\
  }\textbf {\bibinfo {volume} {85}},\ \bibinfo {pages} {1} (\bibinfo {year}
  {2012})}\BibitemShut {NoStop}%
\bibitem [{\citenamefont {Gu{\'{e}}not}\ \emph {et~al.}(2014)\citenamefont
  {Gu{\'{e}}not}, \citenamefont {Kroon}, \citenamefont {Balogh}, \citenamefont
  {Larsen}, \citenamefont {Kotur}, \citenamefont {Miranda}, \citenamefont
  {Fordell}, \citenamefont {Johnsson}, \citenamefont {Mauritsson},
  \citenamefont {Gisselbrecht}, \citenamefont {Varj{\'{u}}}, \citenamefont
  {Arnold}, \citenamefont {Carette}, \citenamefont {Kheifets}, \citenamefont
  {Lindroth}, \citenamefont {Lhuillier},\ and\ \citenamefont
  {Dahlstr{\"{o}}m}}]{Guenot2014}%
  \BibitemOpen
  \bibfield  {author} {\bibinfo {author} {\bibfnamefont {D.}~\bibnamefont
  {Gu{\'{e}}not}}, \bibinfo {author} {\bibfnamefont {D.}~\bibnamefont {Kroon}},
  \bibinfo {author} {\bibfnamefont {E.}~\bibnamefont {Balogh}}, \bibinfo
  {author} {\bibfnamefont {E.~W.}\ \bibnamefont {Larsen}}, \bibinfo {author}
  {\bibfnamefont {M.}~\bibnamefont {Kotur}}, \bibinfo {author} {\bibfnamefont
  {M.}~\bibnamefont {Miranda}}, \bibinfo {author} {\bibfnamefont
  {T.}~\bibnamefont {Fordell}}, \bibinfo {author} {\bibfnamefont
  {P.}~\bibnamefont {Johnsson}}, \bibinfo {author} {\bibfnamefont
  {J.}~\bibnamefont {Mauritsson}}, \bibinfo {author} {\bibfnamefont
  {M.}~\bibnamefont {Gisselbrecht}}, \bibinfo {author} {\bibfnamefont
  {K.}~\bibnamefont {Varj{\'{u}}}}, \bibinfo {author} {\bibfnamefont {C.~L.}\
  \bibnamefont {Arnold}}, \bibinfo {author} {\bibfnamefont {T.}~\bibnamefont
  {Carette}}, \bibinfo {author} {\bibfnamefont {A.~S.}\ \bibnamefont
  {Kheifets}}, \bibinfo {author} {\bibfnamefont {E.}~\bibnamefont {Lindroth}},
  \bibinfo {author} {\bibfnamefont {A.}~\bibnamefont {Lhuillier}},\ and\
  \bibinfo {author} {\bibfnamefont {J.~M.}\ \bibnamefont {Dahlstr{\"{o}}m}},\
  }\bibfield  {title} {\bibinfo {title} {{Measurements of relative
  photoemission time delays in noble gas atoms}},\ }\bibfield  {journal}
  {\bibinfo  {journal} {Journal of Physics B: Atomic, Molecular and Optical
  Physics}\ }\textbf {\bibinfo {volume} {47}},\ \href
  {https://doi.org/10.1088/0953-4075/47/24/245602}
  {10.1088/0953-4075/47/24/245602} (\bibinfo {year} {2014})\BibitemShut
  {NoStop}%
\bibitem [{\citenamefont {Palatchi}\ \emph {et~al.}(2014)\citenamefont
  {Palatchi}, \citenamefont {Dahlstr{\"{o}}m}, \citenamefont {Kheifets},
  \citenamefont {Ivanov}, \citenamefont {Canaday}, \citenamefont {Agostini},\
  and\ \citenamefont {Dimauro}}]{Palatchi2014}%
  \BibitemOpen
  \bibfield  {author} {\bibinfo {author} {\bibfnamefont {C.}~\bibnamefont
  {Palatchi}}, \bibinfo {author} {\bibfnamefont {J.~M.}\ \bibnamefont
  {Dahlstr{\"{o}}m}}, \bibinfo {author} {\bibfnamefont {A.~S.}\ \bibnamefont
  {Kheifets}}, \bibinfo {author} {\bibfnamefont {I.~A.}\ \bibnamefont
  {Ivanov}}, \bibinfo {author} {\bibfnamefont {D.~M.}\ \bibnamefont {Canaday}},
  \bibinfo {author} {\bibfnamefont {P.}~\bibnamefont {Agostini}},\ and\
  \bibinfo {author} {\bibfnamefont {L.~F.}\ \bibnamefont {Dimauro}},\
  }\bibfield  {title} {\bibinfo {title} {{Atomic delay in helium, neon, argon
  and krypton}},\ }\href {https://doi.org/10.1088/0953-4075/47/24/245003}
  {\bibfield  {journal} {\bibinfo  {journal} {Journal of Physics B: Atomic,
  Molecular and Optical Physics}\ }\textbf {\bibinfo {volume} {47}},\ \bibinfo
  {pages} {245003} (\bibinfo {year} {2014})}\BibitemShut {NoStop}%
\bibitem [{\citenamefont {Heuser}\ \emph {et~al.}(2016)\citenamefont {Heuser},
  \citenamefont {Gal{\'{a}}n}, \citenamefont {Cirelli}, \citenamefont
  {Marante}, \citenamefont {Sabbar}, \citenamefont {Boge}, \citenamefont
  {Lucchini}, \citenamefont {Gallmann}, \citenamefont {Ivanov}, \citenamefont
  {Kheifets}, \citenamefont {Dahlstr{\"{o}}m}, \citenamefont {Lindroth},
  \citenamefont {Argenti}, \citenamefont {Mart{\'{i}}n},\ and\ \citenamefont
  {Keller}}]{Heuser2016}%
  \BibitemOpen
  \bibfield  {author} {\bibinfo {author} {\bibfnamefont {S.}~\bibnamefont
  {Heuser}}, \bibinfo {author} {\bibfnamefont {{\'{A}}.~J.}\ \bibnamefont
  {Gal{\'{a}}n}}, \bibinfo {author} {\bibfnamefont {C.}~\bibnamefont
  {Cirelli}}, \bibinfo {author} {\bibfnamefont {C.}~\bibnamefont {Marante}},
  \bibinfo {author} {\bibfnamefont {M.}~\bibnamefont {Sabbar}}, \bibinfo
  {author} {\bibfnamefont {R.}~\bibnamefont {Boge}}, \bibinfo {author}
  {\bibfnamefont {M.}~\bibnamefont {Lucchini}}, \bibinfo {author}
  {\bibfnamefont {L.}~\bibnamefont {Gallmann}}, \bibinfo {author}
  {\bibfnamefont {I.}~\bibnamefont {Ivanov}}, \bibinfo {author} {\bibfnamefont
  {A.~S.}\ \bibnamefont {Kheifets}}, \bibinfo {author} {\bibfnamefont {J.~M.}\
  \bibnamefont {Dahlstr{\"{o}}m}}, \bibinfo {author} {\bibfnamefont
  {E.}~\bibnamefont {Lindroth}}, \bibinfo {author} {\bibfnamefont
  {L.}~\bibnamefont {Argenti}}, \bibinfo {author} {\bibfnamefont
  {F.}~\bibnamefont {Mart{\'{i}}n}},\ and\ \bibinfo {author} {\bibfnamefont
  {U.}~\bibnamefont {Keller}},\ }\bibfield  {title} {\bibinfo {title} {{Angular
  dependence of photoemission time delay in helium}},\ }\href
  {https://doi.org/10.1103/PhysRevA.94.063409} {\bibfield  {journal} {\bibinfo
  {journal} {Physical Review A}\ }\textbf {\bibinfo {volume} {94}},\ \bibinfo
  {pages} {1} (\bibinfo {year} {2016})}\BibitemShut {NoStop}%
\bibitem [{\citenamefont {Huppert}\ \emph {et~al.}(2016)\citenamefont
  {Huppert}, \citenamefont {Jordan}, \citenamefont {Baykusheva}, \citenamefont
  {{Von Conta}},\ and\ \citenamefont {W{\"{o}}rner}}]{Huppert2016a}%
  \BibitemOpen
  \bibfield  {author} {\bibinfo {author} {\bibfnamefont {M.}~\bibnamefont
  {Huppert}}, \bibinfo {author} {\bibfnamefont {I.}~\bibnamefont {Jordan}},
  \bibinfo {author} {\bibfnamefont {D.}~\bibnamefont {Baykusheva}}, \bibinfo
  {author} {\bibfnamefont {A.}~\bibnamefont {{Von Conta}}},\ and\ \bibinfo
  {author} {\bibfnamefont {H.~J.}\ \bibnamefont {W{\"{o}}rner}},\ }\bibfield
  {title} {\bibinfo {title} {{Attosecond Delays in Molecular
  Photoionization}},\ }\href {https://doi.org/10.1103/PhysRevLett.117.093001}
  {\bibfield  {journal} {\bibinfo  {journal} {Physical Review Letters}\
  }\textbf {\bibinfo {volume} {117}},\ \bibinfo {pages} {1} (\bibinfo {year}
  {2016})},\ \Eprint {https://arxiv.org/abs/1607.07435} {arXiv:1607.07435}
  \BibitemShut {NoStop}%
\bibitem [{\citenamefont {Jordan}\ \emph {et~al.}(2017)\citenamefont {Jordan},
  \citenamefont {Huppert}, \citenamefont {Pabst}, \citenamefont {Kheifets},
  \citenamefont {Baykusheva},\ and\ \citenamefont {W{\"{o}}rner}}]{Jordan2017}%
  \BibitemOpen
  \bibfield  {author} {\bibinfo {author} {\bibfnamefont {I.}~\bibnamefont
  {Jordan}}, \bibinfo {author} {\bibfnamefont {M.}~\bibnamefont {Huppert}},
  \bibinfo {author} {\bibfnamefont {S.}~\bibnamefont {Pabst}}, \bibinfo
  {author} {\bibfnamefont {A.~S.}\ \bibnamefont {Kheifets}}, \bibinfo {author}
  {\bibfnamefont {D.}~\bibnamefont {Baykusheva}},\ and\ \bibinfo {author}
  {\bibfnamefont {H.~J.}\ \bibnamefont {W{\"{o}}rner}},\ }\bibfield  {title}
  {\bibinfo {title} {{Spin-orbit delays in photoemission}},\ }\href
  {https://doi.org/10.1103/PhysRevA.95.013404} {\bibfield  {journal} {\bibinfo
  {journal} {Physical Review A}\ }\textbf {\bibinfo {volume} {95}},\ \bibinfo
  {pages} {1} (\bibinfo {year} {2017})}\BibitemShut {NoStop}%
\bibitem [{\citenamefont {Loriot}\ \emph {et~al.}(2017)\citenamefont {Loriot},
  \citenamefont {Marciniak}, \citenamefont {Karras}, \citenamefont {Schindler},
  \citenamefont {Renois-Predelus}, \citenamefont {Compagnon}, \citenamefont
  {Concina}, \citenamefont {Br{\'{e}}dy}, \citenamefont {Celep}, \citenamefont
  {Bordas}, \citenamefont {Constant},\ and\ \citenamefont
  {L{\'{e}}pine}}]{Loriot2017}%
  \BibitemOpen
  \bibfield  {author} {\bibinfo {author} {\bibfnamefont {V.}~\bibnamefont
  {Loriot}}, \bibinfo {author} {\bibfnamefont {A.}~\bibnamefont {Marciniak}},
  \bibinfo {author} {\bibfnamefont {G.}~\bibnamefont {Karras}}, \bibinfo
  {author} {\bibfnamefont {B.}~\bibnamefont {Schindler}}, \bibinfo {author}
  {\bibfnamefont {G.}~\bibnamefont {Renois-Predelus}}, \bibinfo {author}
  {\bibfnamefont {I.}~\bibnamefont {Compagnon}}, \bibinfo {author}
  {\bibfnamefont {B.}~\bibnamefont {Concina}}, \bibinfo {author} {\bibfnamefont
  {R.}~\bibnamefont {Br{\'{e}}dy}}, \bibinfo {author} {\bibfnamefont
  {G.}~\bibnamefont {Celep}}, \bibinfo {author} {\bibfnamefont
  {C.}~\bibnamefont {Bordas}}, \bibinfo {author} {\bibfnamefont
  {E.}~\bibnamefont {Constant}},\ and\ \bibinfo {author} {\bibfnamefont
  {F.}~\bibnamefont {L{\'{e}}pine}},\ }\bibfield  {title} {\bibinfo {title}
  {{Angularly resolved RABBITT using a second harmonic pulse}},\ }\bibfield
  {journal} {\bibinfo  {journal} {Journal of Optics (United Kingdom)}\ }\textbf
  {\bibinfo {volume} {19}},\ \href {https://doi.org/10.1088/2040-8986/aa8e10}
  {10.1088/2040-8986/aa8e10} (\bibinfo {year} {2017})\BibitemShut {NoStop}%
\bibitem [{\citenamefont {Bray}\ \emph {et~al.}(2018)\citenamefont {Bray},
  \citenamefont {Naseem},\ and\ \citenamefont {Kheifets}}]{Bray2018}%
  \BibitemOpen
  \bibfield  {author} {\bibinfo {author} {\bibfnamefont {A.~W.}\ \bibnamefont
  {Bray}}, \bibinfo {author} {\bibfnamefont {F.}~\bibnamefont {Naseem}},\ and\
  \bibinfo {author} {\bibfnamefont {A.~S.}\ \bibnamefont {Kheifets}},\
  }\bibfield  {title} {\bibinfo {title} {{Simulation of angular-resolved
  RABBITT measurements in noble-gas atoms}},\ }\href
  {https://doi.org/10.1103/PhysRevA.97.063404} {\bibfield  {journal} {\bibinfo
  {journal} {Physical Review A}\ }\textbf {\bibinfo {volume} {97}},\ \bibinfo
  {pages} {1} (\bibinfo {year} {2018})}\BibitemShut {NoStop}%
\bibitem [{\citenamefont {Vos}\ \emph {et~al.}(2018)\citenamefont {Vos},
  \citenamefont {Cattaneo}, \citenamefont {Patchkovskii}, \citenamefont
  {Zimmermann}, \citenamefont {Cirelli}, \citenamefont {Lucchini},
  \citenamefont {Kheifets}, \citenamefont {Landsman},\ and\ \citenamefont
  {Keller}}]{Vos2018}%
  \BibitemOpen
  \bibfield  {author} {\bibinfo {author} {\bibfnamefont {J.}~\bibnamefont
  {Vos}}, \bibinfo {author} {\bibfnamefont {L.}~\bibnamefont {Cattaneo}},
  \bibinfo {author} {\bibfnamefont {S.}~\bibnamefont {Patchkovskii}}, \bibinfo
  {author} {\bibfnamefont {T.}~\bibnamefont {Zimmermann}}, \bibinfo {author}
  {\bibfnamefont {C.}~\bibnamefont {Cirelli}}, \bibinfo {author} {\bibfnamefont
  {M.}~\bibnamefont {Lucchini}}, \bibinfo {author} {\bibfnamefont
  {A.}~\bibnamefont {Kheifets}}, \bibinfo {author} {\bibfnamefont {A.~S.}\
  \bibnamefont {Landsman}},\ and\ \bibinfo {author} {\bibfnamefont
  {U.}~\bibnamefont {Keller}},\ }\bibfield  {title} {\bibinfo {title}
  {{Orientation-dependent stereo Wigner time delay and electron localization in
  a small molecule}},\ }\href {https://doi.org/10.1126/science.aao4731}
  {\bibfield  {journal} {\bibinfo  {journal} {Science}\ }\textbf {\bibinfo
  {volume} {360}},\ \bibinfo {pages} {1326} (\bibinfo {year}
  {2018})}\BibitemShut {NoStop}%
\bibitem [{\citenamefont {Goldsmith}\ \emph {et~al.}(2019)\citenamefont
  {Goldsmith}, \citenamefont {Jaro{\'{n}}-Becker}, \citenamefont {Becker},
  \citenamefont {Goldsmith}, \citenamefont {Jaro{\'{n}}-Becker},\ and\
  \citenamefont {Becker}}]{Goldsmith2019}%
  \BibitemOpen
  \bibfield  {author} {\bibinfo {author} {\bibfnamefont {C.}~\bibnamefont
  {Goldsmith}}, \bibinfo {author} {\bibfnamefont {A.}~\bibnamefont
  {Jaro{\'{n}}-Becker}}, \bibinfo {author} {\bibfnamefont {A.}~\bibnamefont
  {Becker}}, \bibinfo {author} {\bibfnamefont {C.}~\bibnamefont {Goldsmith}},
  \bibinfo {author} {\bibfnamefont {A.}~\bibnamefont {Jaro{\'{n}}-Becker}},\
  and\ \bibinfo {author} {\bibfnamefont {A.}~\bibnamefont {Becker}},\
  }\bibfield  {title} {\bibinfo {title} {{Attosecond Streaking Time Delays:
  Finite-Range Interpretation and Applications}},\ }\href
  {https://doi.org/10.3390/app9030492} {\bibfield  {journal} {\bibinfo
  {journal} {Applied Sciences}\ }\textbf {\bibinfo {volume} {9}},\ \bibinfo
  {pages} {492} (\bibinfo {year} {2019})}\BibitemShut {NoStop}%
\bibitem [{\citenamefont {Trebino}\ and\ \citenamefont
  {Kane}(1993)}]{Trebino1993}%
  \BibitemOpen
  \bibfield  {author} {\bibinfo {author} {\bibfnamefont {R.}~\bibnamefont
  {Trebino}}\ and\ \bibinfo {author} {\bibfnamefont {D.~J.}\ \bibnamefont
  {Kane}},\ }\bibfield  {title} {\bibinfo {title} {{Using phase retrieval to
  measure the intensity and phase of ultrashort pulses: frequency-resolved
  optical gating}},\ }\href {https://doi.org/10.1364/josaa.10.001101}
  {\bibfield  {journal} {\bibinfo  {journal} {Journal of the Optical Society of
  America A}\ }\textbf {\bibinfo {volume} {10}},\ \bibinfo {pages} {1101}
  (\bibinfo {year} {1993})}\BibitemShut {NoStop}%
\bibitem [{\citenamefont {Iaconis}\ and\ \citenamefont
  {Walmsley}(1998)}]{Iaconis1998}%
  \BibitemOpen
  \bibfield  {author} {\bibinfo {author} {\bibfnamefont {C.}~\bibnamefont
  {Iaconis}}\ and\ \bibinfo {author} {\bibfnamefont {I.~A.}\ \bibnamefont
  {Walmsley}},\ }\bibfield  {title} {\bibinfo {title} {{Spectral phase
  interferometry for direct electric-field reconstruction of ultrashort optical
  pulses.}},\ }\href {http://www.ncbi.nlm.nih.gov/pubmed/18087344} {\bibfield
  {journal} {\bibinfo  {journal} {Optics letters}\ }\textbf {\bibinfo {volume}
  {23}},\ \bibinfo {pages} {792} (\bibinfo {year} {1998})}\BibitemShut
  {NoStop}%
\bibitem [{\citenamefont {Veniard}\ \emph {et~al.}(1996)\citenamefont
  {Veniard}, \citenamefont {Taieb},\ and\ \citenamefont
  {Maquet}}]{Veniard1996}%
  \BibitemOpen
  \bibfield  {author} {\bibinfo {author} {\bibfnamefont {V.}~\bibnamefont
  {Veniard}}, \bibinfo {author} {\bibfnamefont {R.}~\bibnamefont {Taieb}},\
  and\ \bibinfo {author} {\bibfnamefont {A.}~\bibnamefont {Maquet}},\
  }\bibfield  {title} {\bibinfo {title} {{Phase dependence of N+1 color
  N{\textgreater}1 ir-uv photoionization of atoms with higher harmonics}},\
  }\href@noop {} {\bibfield  {journal} {\bibinfo  {journal} {Physical Review
  A}\ }\textbf {\bibinfo {volume} {54}},\ \bibinfo {pages} {721} (\bibinfo
  {year} {1996})}\BibitemShut {NoStop}%
\bibitem [{\citenamefont {Muller}(2002)}]{Muller2002}%
  \BibitemOpen
  \bibfield  {author} {\bibinfo {author} {\bibfnamefont {H.~G.}\ \bibnamefont
  {Muller}},\ }\bibfield  {title} {\bibinfo {title} {{Reconstruction of
  attosecond harmonic beating by interference of two-photon transitions}},\
  }\href {https://doi.org/10.1007/s00340-002-0894-8} {\bibfield  {journal}
  {\bibinfo  {journal} {Applied Physics B: Lasers and Optics}\ }\textbf
  {\bibinfo {volume} {74}},\ \bibinfo {pages} {17} (\bibinfo {year}
  {2002})}\BibitemShut {NoStop}%
\bibitem [{\citenamefont {Dahlstr{\"{o}}m}\ \emph {et~al.}(2013)\citenamefont
  {Dahlstr{\"{o}}m}, \citenamefont {Gu{\'{e}}not}, \citenamefont
  {Kl{\"{u}}nder}, \citenamefont {Gisselbrecht}, \citenamefont {Mauritsson},
  \citenamefont {L'Huillier}, \citenamefont {Maquet},\ and\ \citenamefont
  {Ta{\"{i}}eb}}]{Dahlstrom2013}%
  \BibitemOpen
  \bibfield  {author} {\bibinfo {author} {\bibfnamefont {J.~M.}\ \bibnamefont
  {Dahlstr{\"{o}}m}}, \bibinfo {author} {\bibfnamefont {D.}~\bibnamefont
  {Gu{\'{e}}not}}, \bibinfo {author} {\bibfnamefont {K.}~\bibnamefont
  {Kl{\"{u}}nder}}, \bibinfo {author} {\bibfnamefont {M.}~\bibnamefont
  {Gisselbrecht}}, \bibinfo {author} {\bibfnamefont {J.}~\bibnamefont
  {Mauritsson}}, \bibinfo {author} {\bibfnamefont {A.}~\bibnamefont
  {L'Huillier}}, \bibinfo {author} {\bibfnamefont {A.}~\bibnamefont {Maquet}},\
  and\ \bibinfo {author} {\bibfnamefont {R.}~\bibnamefont {Ta{\"{i}}eb}},\
  }\bibfield  {title} {\bibinfo {title} {{Theory of attosecond delays in
  laser-assisted photoionization}},\ }\href
  {https://doi.org/10.1016/j.chemphys.2012.01.017} {\bibfield  {journal}
  {\bibinfo  {journal} {Chemical Physics}\ }\textbf {\bibinfo {volume} {414}},\
  \bibinfo {pages} {53} (\bibinfo {year} {2013})},\ \Eprint
  {https://arxiv.org/abs/arXiv:1112.4144v2} {arXiv:arXiv:1112.4144v2}
  \BibitemShut {NoStop}%
\bibitem [{\citenamefont {Baykusheva}\ and\ \citenamefont
  {W{\"{o}}rner}(2017)}]{Baykusheva2017}%
  \BibitemOpen
  \bibfield  {author} {\bibinfo {author} {\bibfnamefont {D.}~\bibnamefont
  {Baykusheva}}\ and\ \bibinfo {author} {\bibfnamefont {H.~J.}\ \bibnamefont
  {W{\"{o}}rner}},\ }\bibfield  {title} {\bibinfo {title} {{Theory of
  Attosecond Delays in Molecular Photoionization}},\ }\bibfield  {journal}
  {\bibinfo  {journal} {Journal of Chemical Physics}\ }\textbf {\bibinfo
  {volume} {146}},\ \href {https://doi.org/10.1103/PhysRevLett.117.093001}
  {10.1103/PhysRevLett.117.093001} (\bibinfo {year} {2017})\BibitemShut
  {NoStop}%
\bibitem [{\citenamefont {Jones}\ \emph {et~al.}(2005)\citenamefont {Jones},
  \citenamefont {Moll}, \citenamefont {Thorpe},\ and\ \citenamefont
  {Ye}}]{Jones2005}%
  \BibitemOpen
  \bibfield  {author} {\bibinfo {author} {\bibfnamefont {R.~J.}\ \bibnamefont
  {Jones}}, \bibinfo {author} {\bibfnamefont {K.~D.}\ \bibnamefont {Moll}},
  \bibinfo {author} {\bibfnamefont {M.~J.}\ \bibnamefont {Thorpe}},\ and\
  \bibinfo {author} {\bibfnamefont {J.}~\bibnamefont {Ye}},\ }\bibfield
  {title} {\bibinfo {title} {{Phase-coherent frequency combs in the vacuum
  ultraviolet via high-harmonic generation inside a femtosecond enhancement
  cavity}},\ }\href {https://doi.org/10.1103/PhysRevLett.94.193201} {\bibfield
  {journal} {\bibinfo  {journal} {Physical Review Letters}\ }\textbf {\bibinfo
  {volume} {94}},\ \bibinfo {pages} {1} (\bibinfo {year} {2005})}\BibitemShut
  {NoStop}%
\bibitem [{\citenamefont {Pupeza}\ \emph {et~al.}(2013)\citenamefont {Pupeza},
  \citenamefont {Holzberger}, \citenamefont {Eidam}, \citenamefont {Carstens},
  \citenamefont {Esser}, \citenamefont {Weitenberg}, \citenamefont
  {Ru{\ss}b{\"{u}}ldt}, \citenamefont {Rauschenberger}, \citenamefont
  {Limpert}, \citenamefont {Udem}, \citenamefont {T{\"{u}}nnermann},
  \citenamefont {H{\"{a}}nsch}, \citenamefont {Apolonski}, \citenamefont
  {Krausz},\ and\ \citenamefont {Fill}}]{Pupeza2013}%
  \BibitemOpen
  \bibfield  {author} {\bibinfo {author} {\bibfnamefont {I.}~\bibnamefont
  {Pupeza}}, \bibinfo {author} {\bibfnamefont {S.}~\bibnamefont {Holzberger}},
  \bibinfo {author} {\bibfnamefont {T.}~\bibnamefont {Eidam}}, \bibinfo
  {author} {\bibfnamefont {H.}~\bibnamefont {Carstens}}, \bibinfo {author}
  {\bibfnamefont {D.}~\bibnamefont {Esser}}, \bibinfo {author} {\bibfnamefont
  {J.}~\bibnamefont {Weitenberg}}, \bibinfo {author} {\bibfnamefont
  {P.}~\bibnamefont {Ru{\ss}b{\"{u}}ldt}}, \bibinfo {author} {\bibfnamefont
  {J.}~\bibnamefont {Rauschenberger}}, \bibinfo {author} {\bibfnamefont
  {J.}~\bibnamefont {Limpert}}, \bibinfo {author} {\bibfnamefont
  {T.}~\bibnamefont {Udem}}, \bibinfo {author} {\bibfnamefont {A.}~\bibnamefont
  {T{\"{u}}nnermann}}, \bibinfo {author} {\bibfnamefont {T.~W.}\ \bibnamefont
  {H{\"{a}}nsch}}, \bibinfo {author} {\bibfnamefont {A.}~\bibnamefont
  {Apolonski}}, \bibinfo {author} {\bibfnamefont {F.}~\bibnamefont {Krausz}},\
  and\ \bibinfo {author} {\bibfnamefont {E.}~\bibnamefont {Fill}},\ }\bibfield
  {title} {\bibinfo {title} {{Compact high-repetition-rate source of coherent
  100 eV radiation}},\ }\href {https://doi.org/10.1038/nphoton.2013.156}
  {\bibfield  {journal} {\bibinfo  {journal} {Nature Photonics}\ }\textbf
  {\bibinfo {volume} {7}},\ \bibinfo {pages} {608} (\bibinfo {year}
  {2013})}\BibitemShut {NoStop}%
\bibitem [{\citenamefont {Carstens}\ \emph {et~al.}(2016)\citenamefont
  {Carstens}, \citenamefont {Guggenmos}, \citenamefont {Pupeza}, \citenamefont
  {Esser}, \citenamefont {Saule}, \citenamefont {Holzberger}, \citenamefont
  {Kleineberg}, \citenamefont {Tosa}, \citenamefont {Eidam}, \citenamefont
  {Pervak}, \citenamefont {Krausz}, \citenamefont {Limpert}, \citenamefont
  {Jocher}, \citenamefont {Lilienfein}, \citenamefont {T{\"{u}}nnermann},\ and\
  \citenamefont {H{\"{o}}gner}}]{Carstens2016}%
  \BibitemOpen
  \bibfield  {author} {\bibinfo {author} {\bibfnamefont {H.}~\bibnamefont
  {Carstens}}, \bibinfo {author} {\bibfnamefont {A.}~\bibnamefont {Guggenmos}},
  \bibinfo {author} {\bibfnamefont {I.}~\bibnamefont {Pupeza}}, \bibinfo
  {author} {\bibfnamefont {D.}~\bibnamefont {Esser}}, \bibinfo {author}
  {\bibfnamefont {T.}~\bibnamefont {Saule}}, \bibinfo {author} {\bibfnamefont
  {S.}~\bibnamefont {Holzberger}}, \bibinfo {author} {\bibfnamefont
  {U.}~\bibnamefont {Kleineberg}}, \bibinfo {author} {\bibfnamefont
  {V.}~\bibnamefont {Tosa}}, \bibinfo {author} {\bibfnamefont {T.}~\bibnamefont
  {Eidam}}, \bibinfo {author} {\bibfnamefont {V.}~\bibnamefont {Pervak}},
  \bibinfo {author} {\bibfnamefont {F.}~\bibnamefont {Krausz}}, \bibinfo
  {author} {\bibfnamefont {J.}~\bibnamefont {Limpert}}, \bibinfo {author}
  {\bibfnamefont {C.}~\bibnamefont {Jocher}}, \bibinfo {author} {\bibfnamefont
  {N.}~\bibnamefont {Lilienfein}}, \bibinfo {author} {\bibfnamefont
  {A.}~\bibnamefont {T{\"{u}}nnermann}},\ and\ \bibinfo {author} {\bibfnamefont
  {M.}~\bibnamefont {H{\"{o}}gner}},\ }\bibfield  {title} {\bibinfo {title}
  {{High-harmonic generation at 250 MHz with photon energies exceeding 100
  eV}},\ }\href {https://doi.org/10.1364/optica.3.000366} {\bibfield  {journal}
  {\bibinfo  {journal} {Optica}\ }\textbf {\bibinfo {volume} {3}},\ \bibinfo
  {pages} {366} (\bibinfo {year} {2016})}\BibitemShut {NoStop}%
\bibitem [{\citenamefont {Higuet}\ \emph {et~al.}(2009)\citenamefont {Higuet},
  \citenamefont {M{\'{e}}vel}, \citenamefont {Mairesse}, \citenamefont
  {Boullet}, \citenamefont {Limpert}, \citenamefont {Petit}, \citenamefont
  {Cormier}, \citenamefont {Constant}, \citenamefont {Fabre},\ and\
  \citenamefont {Zaouter}}]{Higuet2009}%
  \BibitemOpen
  \bibfield  {author} {\bibinfo {author} {\bibfnamefont {J.}~\bibnamefont
  {Higuet}}, \bibinfo {author} {\bibfnamefont {E.}~\bibnamefont {M{\'{e}}vel}},
  \bibinfo {author} {\bibfnamefont {Y.}~\bibnamefont {Mairesse}}, \bibinfo
  {author} {\bibfnamefont {J.}~\bibnamefont {Boullet}}, \bibinfo {author}
  {\bibfnamefont {J.}~\bibnamefont {Limpert}}, \bibinfo {author} {\bibfnamefont
  {S.}~\bibnamefont {Petit}}, \bibinfo {author} {\bibfnamefont
  {E.}~\bibnamefont {Cormier}}, \bibinfo {author} {\bibfnamefont
  {E.}~\bibnamefont {Constant}}, \bibinfo {author} {\bibfnamefont
  {B.}~\bibnamefont {Fabre}},\ and\ \bibinfo {author} {\bibfnamefont
  {Y.}~\bibnamefont {Zaouter}},\ }\bibfield  {title} {\bibinfo {title}
  {{High-order harmonic generation at a megahertz-level repetition rate
  directly driven by an ytterbium-doped-fiber chirped-pulse amplification
  system}},\ }\href {https://doi.org/10.1364/ol.34.001489} {\bibfield
  {journal} {\bibinfo  {journal} {Optics Letters}\ }\textbf {\bibinfo {volume}
  {34}},\ \bibinfo {pages} {1489} (\bibinfo {year} {2009})}\BibitemShut
  {NoStop}%
\bibitem [{\citenamefont {Russbueldt}\ \emph {et~al.}(2011)\citenamefont
  {Russbueldt}, \citenamefont {H{\"{a}}nsch}, \citenamefont {Weitenberg},
  \citenamefont {Krausz}, \citenamefont {Kling}, \citenamefont {Udem},
  \citenamefont {Sartorius}, \citenamefont {Schneider}, \citenamefont
  {Vernaleken}, \citenamefont {Stebbings}, \citenamefont {Poprawe},
  \citenamefont {Hoffmann},\ and\ \citenamefont {Hommelhoff}}]{Russbueldt2011}%
  \BibitemOpen
  \bibfield  {author} {\bibinfo {author} {\bibfnamefont {P.}~\bibnamefont
  {Russbueldt}}, \bibinfo {author} {\bibfnamefont {T.~W.}\ \bibnamefont
  {H{\"{a}}nsch}}, \bibinfo {author} {\bibfnamefont {J.}~\bibnamefont
  {Weitenberg}}, \bibinfo {author} {\bibfnamefont {F.}~\bibnamefont {Krausz}},
  \bibinfo {author} {\bibfnamefont {M.~F.}\ \bibnamefont {Kling}}, \bibinfo
  {author} {\bibfnamefont {T.}~\bibnamefont {Udem}}, \bibinfo {author}
  {\bibfnamefont {T.}~\bibnamefont {Sartorius}}, \bibinfo {author}
  {\bibfnamefont {W.}~\bibnamefont {Schneider}}, \bibinfo {author}
  {\bibfnamefont {A.}~\bibnamefont {Vernaleken}}, \bibinfo {author}
  {\bibfnamefont {S.~L.}\ \bibnamefont {Stebbings}}, \bibinfo {author}
  {\bibfnamefont {R.}~\bibnamefont {Poprawe}}, \bibinfo {author} {\bibfnamefont
  {H.-D.}\ \bibnamefont {Hoffmann}},\ and\ \bibinfo {author} {\bibfnamefont
  {P.}~\bibnamefont {Hommelhoff}},\ }\bibfield  {title} {\bibinfo {title}
  {{Single-pass high-harmonic generation at 208 MHz repetition rate}},\ }\href
  {https://doi.org/10.1364/ol.36.003428} {\bibfield  {journal} {\bibinfo
  {journal} {Optics Letters}\ }\textbf {\bibinfo {volume} {36}},\ \bibinfo
  {pages} {3428} (\bibinfo {year} {2011})}\BibitemShut {NoStop}%
\bibitem [{\citenamefont {Rothhardt}\ \emph {et~al.}(2014)\citenamefont
  {Rothhardt}, \citenamefont {H{\"{a}}drich}, \citenamefont {Klenke},
  \citenamefont {Demmler}, \citenamefont {Hoffmann}, \citenamefont {Gotschall},
  \citenamefont {Eidam}, \citenamefont {Krebs}, \citenamefont {Limpert},\ and\
  \citenamefont {T{\"{u}}nnermann}}]{Rothhardt2014}%
  \BibitemOpen
  \bibfield  {author} {\bibinfo {author} {\bibfnamefont {J.}~\bibnamefont
  {Rothhardt}}, \bibinfo {author} {\bibfnamefont {S.}~\bibnamefont
  {H{\"{a}}drich}}, \bibinfo {author} {\bibfnamefont {A.}~\bibnamefont
  {Klenke}}, \bibinfo {author} {\bibfnamefont {S.}~\bibnamefont {Demmler}},
  \bibinfo {author} {\bibfnamefont {A.}~\bibnamefont {Hoffmann}}, \bibinfo
  {author} {\bibfnamefont {T.}~\bibnamefont {Gotschall}}, \bibinfo {author}
  {\bibfnamefont {T.}~\bibnamefont {Eidam}}, \bibinfo {author} {\bibfnamefont
  {M.}~\bibnamefont {Krebs}}, \bibinfo {author} {\bibfnamefont
  {J.}~\bibnamefont {Limpert}},\ and\ \bibinfo {author} {\bibfnamefont
  {A.}~\bibnamefont {T{\"{u}}nnermann}},\ }\bibfield  {title} {\bibinfo {title}
  {{53 W average power few-cycle fiber laser system generating soft x rays up
  to the water window}},\ }\href {https://doi.org/10.1364/ol.39.005224}
  {\bibfield  {journal} {\bibinfo  {journal} {Optics Letters}\ }\textbf
  {\bibinfo {volume} {39}},\ \bibinfo {pages} {5224} (\bibinfo {year}
  {2014})}\BibitemShut {NoStop}%
\bibitem [{\citenamefont {Harth}\ \emph {et~al.}(2017)\citenamefont {Harth},
  \citenamefont {Guo}, \citenamefont {Cheng}, \citenamefont {Losquin},
  \citenamefont {Miranda}, \citenamefont {Mikaelsson}, \citenamefont {Heyl},
  \citenamefont {Prochnow}, \citenamefont {Ahrens}, \citenamefont {Morgner},
  \citenamefont {L'Huillier},\ and\ \citenamefont {Arnold}}]{Harth2017}%
  \BibitemOpen
  \bibfield  {author} {\bibinfo {author} {\bibfnamefont {A.}~\bibnamefont
  {Harth}}, \bibinfo {author} {\bibfnamefont {C.}~\bibnamefont {Guo}}, \bibinfo
  {author} {\bibfnamefont {Y.-C.}\ \bibnamefont {Cheng}}, \bibinfo {author}
  {\bibfnamefont {A.}~\bibnamefont {Losquin}}, \bibinfo {author} {\bibfnamefont
  {M.}~\bibnamefont {Miranda}}, \bibinfo {author} {\bibfnamefont
  {S.}~\bibnamefont {Mikaelsson}}, \bibinfo {author} {\bibfnamefont
  {C.}~\bibnamefont {Heyl}}, \bibinfo {author} {\bibfnamefont {O.}~\bibnamefont
  {Prochnow}}, \bibinfo {author} {\bibfnamefont {J.}~\bibnamefont {Ahrens}},
  \bibinfo {author} {\bibfnamefont {U.}~\bibnamefont {Morgner}}, \bibinfo
  {author} {\bibfnamefont {A.}~\bibnamefont {L'Huillier}},\ and\ \bibinfo
  {author} {\bibfnamefont {C.~L.}\ \bibnamefont {Arnold}},\ }\bibfield  {title}
  {\bibinfo {title} {{Compact 200kHz HHG source driven by a few-cycle OPCPA}},\
  }\bibfield  {journal} {\bibinfo  {journal} {Journal of Optics}\ }\href
  {https://doi.org/10.1088/2040-8986/aa9b04} {10.1088/2040-8986/aa9b04}
  (\bibinfo {year} {2017})\BibitemShut {NoStop}%
\bibitem [{\citenamefont {K{\"{u}}hn}\ \emph {et~al.}(2017)\citenamefont
  {K{\"{u}}hn}, \citenamefont {Dumergue}, \citenamefont {Kahaly}, \citenamefont
  {Mondal}, \citenamefont {Calegari}, \citenamefont {Sansone}, \citenamefont
  {Stagira}, \citenamefont {H{\"{a}}drich}, \citenamefont {Rothhardt},
  \citenamefont {Krebs}, \citenamefont {Reid}, \citenamefont {Heyl},\ and\
  \citenamefont {Thomson}}]{Kuhn2017}%
  \BibitemOpen
  \bibfield  {author} {\bibinfo {author} {\bibfnamefont {S.}~\bibnamefont
  {K{\"{u}}hn}}, \bibinfo {author} {\bibfnamefont {M.}~\bibnamefont
  {Dumergue}}, \bibinfo {author} {\bibfnamefont {S.}~\bibnamefont {Kahaly}},
  \bibinfo {author} {\bibfnamefont {S.}~\bibnamefont {Mondal}}, \bibinfo
  {author} {\bibfnamefont {F.}~\bibnamefont {Calegari}}, \bibinfo {author}
  {\bibfnamefont {G.}~\bibnamefont {Sansone}}, \bibinfo {author} {\bibfnamefont
  {S.}~\bibnamefont {Stagira}}, \bibinfo {author} {\bibfnamefont
  {S.}~\bibnamefont {H{\"{a}}drich}}, \bibinfo {author} {\bibfnamefont
  {J.}~\bibnamefont {Rothhardt}}, \bibinfo {author} {\bibfnamefont
  {M.}~\bibnamefont {Krebs}}, \bibinfo {author} {\bibfnamefont {D.~T.}\
  \bibnamefont {Reid}}, \bibinfo {author} {\bibfnamefont {C.~M.}\ \bibnamefont
  {Heyl}},\ and\ \bibinfo {author} {\bibfnamefont {R.~R.}\ \bibnamefont
  {Thomson}},\ }\bibfield  {title} {\bibinfo {title} {{The ELI-ALPS facility :
  the next generation of attosecond sources}},\ }\href@noop {} {\bibfield
  {journal} {\bibinfo  {journal} {Journal of Physics B: Atomic, Molecular and
  Optical Physics}\ }\textbf {\bibinfo {volume} {50}} (\bibinfo {year}
  {2017})}\BibitemShut {NoStop}%
\bibitem [{\citenamefont {Swoboda}\ \emph {et~al.}(2009)\citenamefont
  {Swoboda}, \citenamefont {Dahlstrom}, \citenamefont {Ruchon}, \citenamefont
  {Johnsson}, \citenamefont {Mauritsson}, \citenamefont {Schafer},\ and\
  \citenamefont {L'Huillier}}]{Swoboda2009}%
  \BibitemOpen
  \bibfield  {author} {\bibinfo {author} {\bibfnamefont {M.}~\bibnamefont
  {Swoboda}}, \bibinfo {author} {\bibfnamefont {J.~M.}\ \bibnamefont
  {Dahlstrom}}, \bibinfo {author} {\bibfnamefont {T.}~\bibnamefont {Ruchon}},
  \bibinfo {author} {\bibfnamefont {P.}~\bibnamefont {Johnsson}}, \bibinfo
  {author} {\bibfnamefont {J.}~\bibnamefont {Mauritsson}}, \bibinfo {author}
  {\bibfnamefont {K.~J.}\ \bibnamefont {Schafer}},\ and\ \bibinfo {author}
  {\bibfnamefont {A.}~\bibnamefont {L'Huillier}},\ }\bibfield  {title}
  {\bibinfo {title} {{Intensity Dependence of Laser-Assisted Attosecond
  Photoionization Spectra}},\ }\href
  {https://doi.org/10.1134/S1054660X09150390} {\bibfield  {journal} {\bibinfo
  {journal} {Laser Physics}\ }\textbf {\bibinfo {volume} {19}},\ \bibinfo
  {pages} {1591} (\bibinfo {year} {2009})},\ \Eprint
  {https://arxiv.org/abs/0906.1158} {arXiv:0906.1158} \BibitemShut {NoStop}%
\bibitem [{\citenamefont {Gaumnitz}\ \emph {et~al.}(2018)\citenamefont
  {Gaumnitz}, \citenamefont {Jain},\ and\ \citenamefont
  {W{\"{o}}rner}}]{Gaumnitz2018}%
  \BibitemOpen
  \bibfield  {author} {\bibinfo {author} {\bibfnamefont {T.}~\bibnamefont
  {Gaumnitz}}, \bibinfo {author} {\bibfnamefont {A.}~\bibnamefont {Jain}},\
  and\ \bibinfo {author} {\bibfnamefont {H.~J.}\ \bibnamefont {W{\"{o}}rner}},\
  }\bibfield  {title} {\bibinfo {title} {{Extreme-ultraviolet high-order
  harmonic generation from few-cycle annular beams}},\ }\href
  {https://doi.org/10.1364/ol.43.004506} {\bibfield  {journal} {\bibinfo
  {journal} {Optics Letters}\ }\textbf {\bibinfo {volume} {43}},\ \bibinfo
  {pages} {4506} (\bibinfo {year} {2018})}\BibitemShut {NoStop}%
\bibitem [{\citenamefont {Becker}(1996)}]{Becker1996}%
  \BibitemOpen
  \bibfield  {author} {\bibinfo {author} {\bibfnamefont {D.}~\bibnamefont
  {Becker}, \bibfnamefont {U.~Shirley}},\ }\href@noop {} {\emph {\bibinfo
  {title} {VUV and Soft X-Ray Photoionization}}},\ edited by\ \bibinfo {editor}
  {\bibfnamefont {D.}~\bibnamefont {Becker}, \bibfnamefont {U.~Shirley}}\
  (\bibinfo  {publisher} {Plenum Press},\ \bibinfo {year} {1996})\BibitemShut
  {NoStop}%
\end{thebibliography}%

\end{document}